\documentclass[prl,twocolumn,aps]{revtex4}

\usepackage{graphicx}

\newcommand{\ben}{\begin{eqnarray}}
\newcommand{\een}{\end{eqnarray}}
\newcommand{\la}{\label}

\usepackage{amsmath,amssymb}
\usepackage{eucal}
\usepackage{graphicx}

\begin{document}

\title{On gravitational waves from classical three body problem}

\author{Plamen P Fiziev}

\affiliation{Sofia University Foundation for Theoretical and Computational Physics and Astrophysics, Boulevard
5 James Bourchier, Sofia 1164, Bulgaria \\
{\rm and} BLTF, JINR, Dubna, 141980 Moscow Region, Rusia
}
\date{\today}

\begin{abstract}
Using an effective one body approach we describe in detail gravitational waves from classical three body problem
on a non-rotating straight line and derive their basic physical characteristics.
Special attention is paid to the irregular motions of such systems and to the significance of  double and triple collisions.
The conclusive role of the collinear solutions is also discussed in short. It is shown that the residuals may contain
information about irregular motion of the source of gravitational waves. 
\pacs{}

\end{abstract}
\keywords{gravitational waves, classical three body problem, double collision, triple collision}

\maketitle

\section{Introduction}
Owing to LIGO discoveries \cite{LIGO,LOGOOPEN} we already live in the epoch of the gravitational waves astronomy.
It becomes of critical importance to have a full list of the significant sources of Gravitational Waves (GW)
in the Nature. A description of many of them can be found in \cite{Thorne87,Maggiore,Buonanno15,Miller16b}.
 The commonly recognized astrophysical sources of GW are the inspiraling and merging binary compact objects \cite{LIGO,LOGOOPEN}.
\begin{figure}[!ht]
\centering
\begin{minipage}{8.cm}
\vskip .truecm
\hskip .truecm
\includegraphics[width=.8\textwidth,natwidth=200,natheight=200]{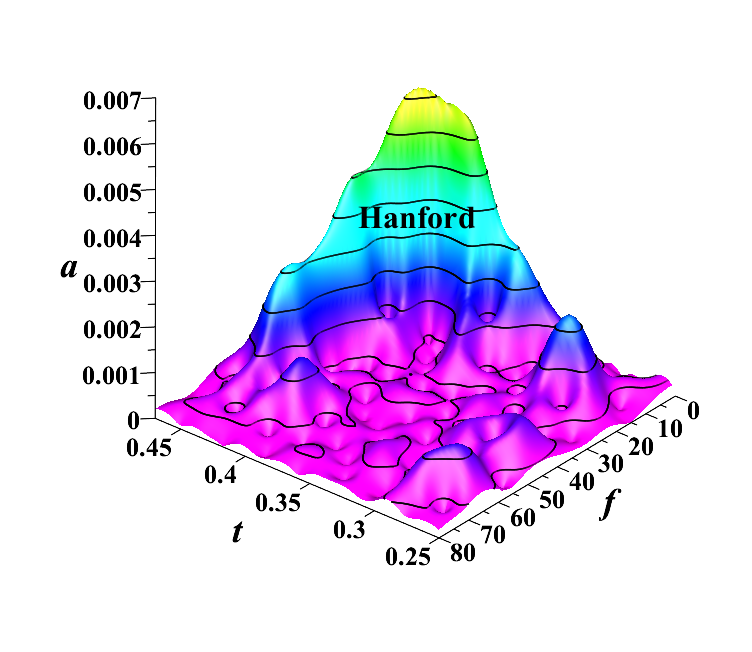}
\vskip .truecm
\hskip .truecm
\includegraphics[width=.7\textwidth,natwidth=200,natheight=200]{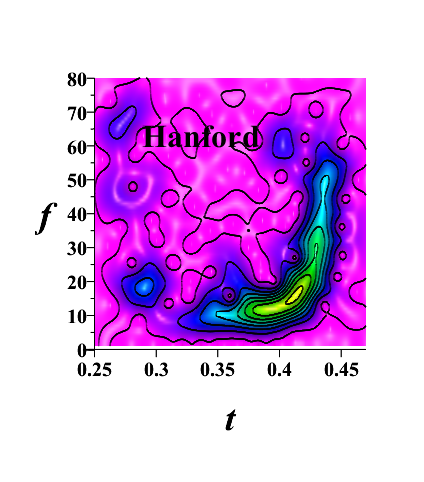}
\vskip .truecm
\hskip .truecm
\includegraphics[width=.9\textwidth,natwidth=200,natheight=200]{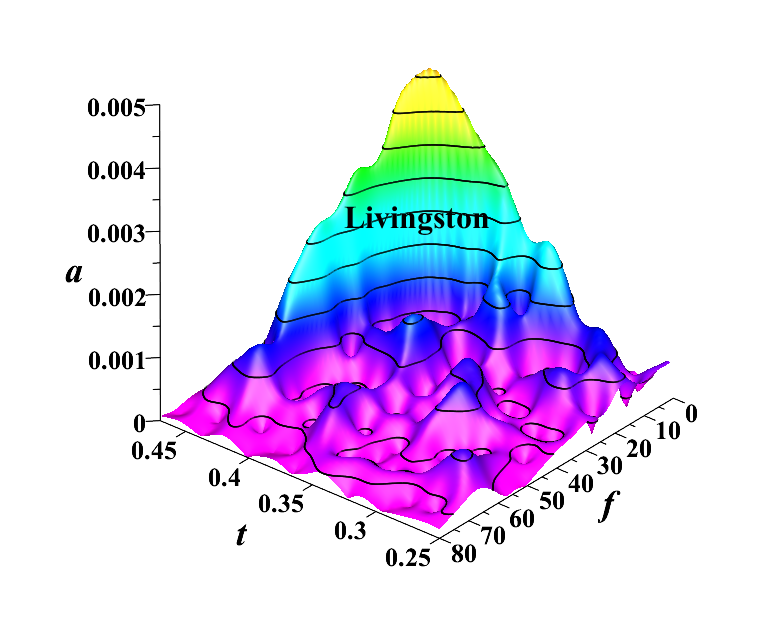}
\vskip .truecm
\hskip .truecm
\includegraphics[width=.65\textwidth,natwidth=200,natheight=200]{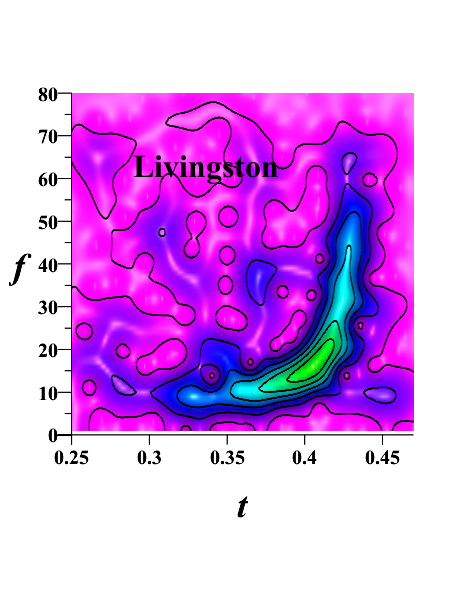}
\end{minipage}
\vskip -.8truecm
\caption{\small The Gabor transform of GW150914 data from \cite{LIGO,LOGOOPEN}.
One sees clear indications of an irregular motion of the source.
(In all figures $f=\text{frequency}$, $a=\text{amplitude}$, $t=\text{time}$.)}
\label{Fig1}
\end{figure}
\begin{figure}[!ht]
\centering
\begin{minipage}{8.cm}
\vskip .truecm
\hskip .truecm
\includegraphics[width=\textwidth,natwidth=200,natheight=200]{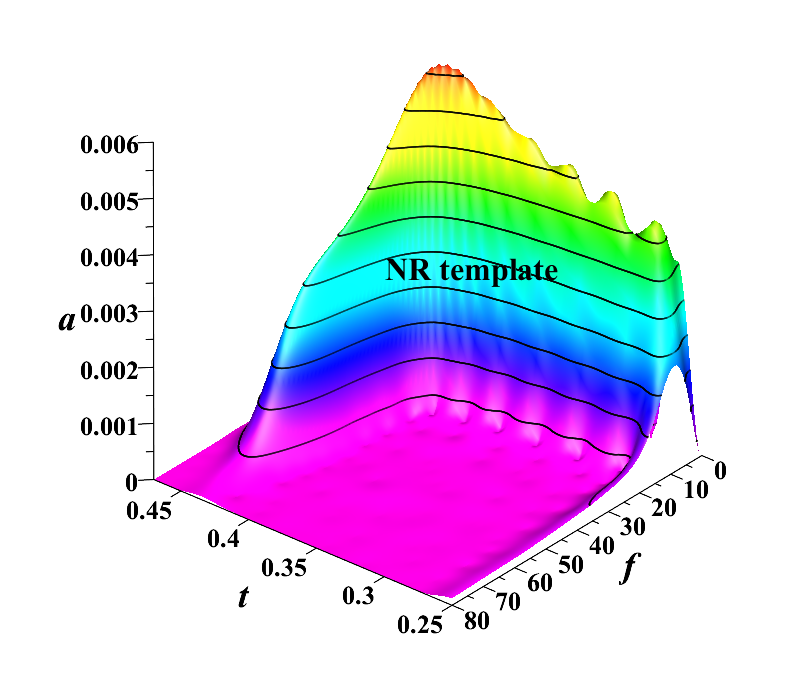}
\vskip .truecm
\hskip .truecm
\includegraphics[width=\textwidth,natwidth=200,natheight=200]{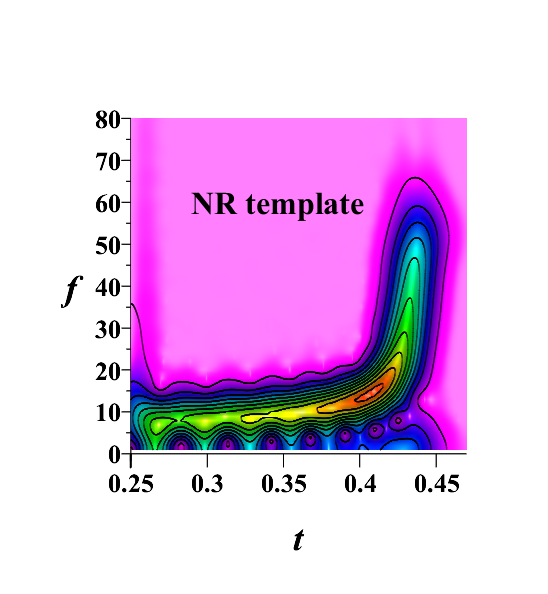}
\end{minipage}
\vskip .truecm
\caption{\small The Gabor transform of NR-template used in \cite{LIGO,LOGOOPEN}.
Here we have no low hills, interpreted entirely as random noise in \cite{LIGO}.}
\label{Fig2}
\end{figure}
A basic condition for a significant radiation of GW by a system of bodies is the large acceleration of at least one of them.
The initial studies of regular periodic solutions of classical Three Body Systems (3BS)
showed that these are not perspective sources of GW for LIGO-type of GW-detectors \cite{Chiba07,Torigoe09,Asada09}
just because large accelerations do not exist,
even for quite complicated 3BS motions \cite{Zeigel71,Arnold06}.

The Letter [12] looks for 3BS periodic sources of GW and emphasizes the role
of the two-body-collisions (2BC) for emission of intense quadrupolar GW.

An irregular motion (deterministic chaos) is impossible in a binary system, see Fig.\ref{Fig2}.
Irregular motions with large accelerations are possible only in N-Body System (NBS) if $N>2$
and only when some singular point is approached.

The experience from the study of binary systems is not much helpful for $N>2$.
For example, in point particle idealization of NBS the energy conservation is not an obstacle for a resonant transfer of an unbounded
amount of energy to one of the particles using the "infinite" Newton reservoir of potential energy of some pair of other particles.
To obtain intense GW, one must consider trajectories which approach singular points and transform a significant amount of
potential energy into kinetic energy.

In the simplest case of an irregular motion $N=3$. Then only two types of singular points exist:  2BC and Three-Body-Collisions (3BC).
In the 18-dimensional set of all 3BS-orbits the set of orbits with 2BC has 16 dimensions
and consists of three analytic manifolds  \cite{Zeigel71,Arnold06}.

In the general case, the set of orbits with 3BC has an unknown structure and analytical properties.
The 3BC in the collinear 3BS was described in \cite{McGehee74}.

The situation may be very complicated.
According to the Poincar\'e hypothesis, in an arbitrary small vicinity
of each point of the phase space of 3BS there may exist all 7 types of qualitatively different solutions of 3BS-dynamics
described by the Chazy classification  \cite{Zeigel71,Arnold06}.

However, dimensional arguments show that 2BC are more probable than 3BC.
Indeed, 2BC can happen for any value of the total angular momentum $\mathbf{K}$ of 3BS.
In contrast, 3BC are possible only if $\mathbf{K}\equiv 0$, i.e. the corresponding orbits form some unknown nontrivial subset
of this 16-dimensional analytic manifold.

When $\mathbf{K}\equiv 0$, 3BS approaches 3BC
via central configurations of two types:
a) equilateral triangle configuration;
b) collinear configuration.
We do not consider here the case a) in which 2BC are impossible around 3BC.

Thus, our main topic becomes the 3BS on a non-rotating straight line.
It describes the basic state of 3BS and introduces the new special functions
needed to solve the general three body problem \cite{Fiziev87a}.
It is important to stress
that this case is produced just by a special choice of the initial conditions for the general 3-dimensional 3BS.

From a physical point of view we have to know the intensity of GW
emitted not just from the exact solutions with 2BC and 3BC.
Actually, we need to study part of the phase flow which is able to follow closely these
solutions for a long enough time.
It is likely that the needed part of the phase flow has a positive measure in the 18-dimensional manifold of 3BS solutions.
Special analytical and numerical methods must be developed for its study.

Hence, one can consider the collinear solutions of 3BS with 2BC and 3BC
as the simplest physical idealization of the real problem.
In this Letter we make the first step to study the collinear 3BS in the context of GW emission.

\section{Effective one body approach to the three body collinear problem}

Our consideration is based on the introduction of specific hyper-spherical coordinates for collinear 3BS
 on the axis $Ox$ ($O$ being the 3BS-center-of-mass) which were described and used in \cite{Fiziev87a,Fiziev87b,Fiziev86,Fiziev87c}.
The effective one body has a reduced mass $\mu=\sqrt{m_1 m_2 m_3/(m_1+ m_2 +m_3)}$ and kinetic energy
$T={\frac 1 {2\mu}}\left( p_\rho^2+\rho^{-2} p_\varphi^2 \right)$.
A special property of our choice of hyper-spherical coordinates is that
in the absence of interaction between the point masses $m_{1,2,3}$ with the coordinates $x_i=x_i(\rho,\varphi)$
the effective mass $\mu$ moves as a free particle in an effective  two dimensional plane with the polar coordinates $\rho,\varphi$ and origin $O$,
see for details  \cite{Fiziev87a,Fiziev87b,Fiziev86,Fiziev87c}.

\begin{figure}[!ht]
\centering
\begin{minipage}{8.cm}
\vskip .truecm
\hskip .truecm
\includegraphics[width=0.7\textwidth,natwidth=200,natheight=200]{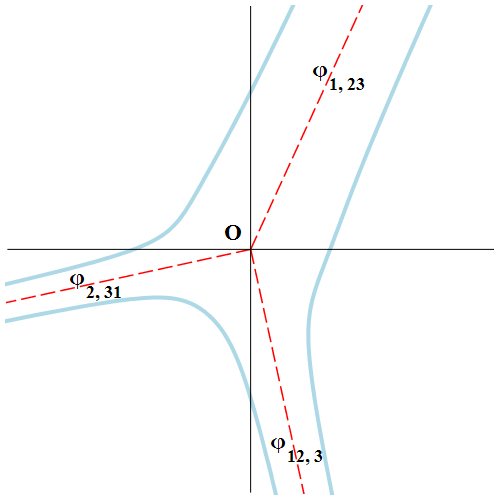}
\vskip .truecm
\caption{\small The movement of an effective body with $E<0$ takes place in three joined channels.
The lines of 2BC are dashed. The comma in $\varphi_{i,jk}$ shows the position of the origin $O$ with respect to the mergered particles $\{jk\}$.}
\label{Fig3}
\end{minipage}
\end{figure}

The most remarkable result is the form of the Newton potential of the collinear 3BS:
\ben
V(\rho,\varphi)=-G\,{\frac {\mu M \alpha(\varphi)} {\rho} },
\la{V}
\een
where $G$ is the Newton constant, $M=m_1+ m_2 +m_3$ and
$\alpha(\varphi)=\sum_{i=1}^3 c_i s_i^{-2} |sin(\varphi -\varphi_{jk,i})|^{-1},\,i\neq j\neq k$
\footnote{The constant quantities $c_i=\cos\psi_i, s_i=\sin\psi_i$, $\psi_i=\arctan(m_i/\mu)$
$\varphi_{23,1}={\frac \pi 6}-{\frac {\psi_2-\psi_3} 2}$,
$\varphi_{3,12}=3{\frac \pi 6}-{\frac{\psi_1-\psi_2} 2}$,
$\varphi_{31,2}=5{\frac \pi 6}-{\frac{\psi_3-\psi_1} 2}$,
are defined by masses $m_i$.}.
The 2BC angles $\varphi_{i,jk}=\varphi_{jk,i}+\pi$ define the directions of 2BS
in the plane with polar coordinates $\{\rho, \varphi\}$, see Fig.\ref{Fig3}.

The equations of motion for hamiltonian $H=T+V$ and their solutions were studied in detail in \cite{Fiziev87a,Fiziev87b,Fiziev86,Fiziev87c}.
For $E=0$ and $E=\infty$ we have two different integrable cases.

Due to the scaling properties of $H$, all other cases, being non-integrable,
are reduced to the cases $E=1$ or $E=-1$ \footnote{The 3BS with potential \eqref{V} has a hidden scaling invariance
under changes $\rho=\bar\rho/\varrho$, $E=\varrho\bar E$, $t=\varrho^{-3/2}\bar t$, $p_\rho=\varrho^{1/2}\bar p_\rho$,
$p_\varphi=\varrho^{-1/2}\bar p_\varphi$, $\nu=\varrho^{3/2}\bar\nu$.
Here $\varrho>0$ is an arbitrary positive constant and $\nu$ is some frequency. The variable $\varphi$ is invariant.
Other nontrivial invariants can be  constructed in an obvious way.
These transformations map every solution of 3BS onto another its solution.},
and demonstrate a more or less irregular motion.
Most of the solutions have random in number series of 2BC and may approach many times 3BC.

We introduce a regularization parameter for all 2BC and suppose that they present ideal elastic strikes
\footnote{This idealization can be easily removed.}.

The 3BC is a saddle-nod singular point in 3-dimen-sional iso-energetic surface. The 3BC happens for $\rho=0$.

Thus, a remarkable property of the hyperspherical coordinates is
a total separation of 2BC and 3BC. Responsible coordinates are $\varphi$ and $\rho$, respectively.

In a small vicinity of the 3BC-point the phase flow of the collinear 3BS splits into four three dimensional subflows with a different behaviour.
Parts of these subflows can go arbitrarily close to 3BC but never reach it.

The real submanifold of the 3BC-trajectories is two dimensional.
On it, the 3BC-trajectories form a node and can be regularized in a similar,
but non-analytical way, as the 2BC-trajectories do,
i.e., as continuous limits of the corresponding parts of the 3 dimensional subflows.
This picture clarifies the results of \cite{McGehee74} and
the wide spread statement that in the vicinity of 3BC
the total phase flow of 3BS cannot be regularized and extended.

In the collinear 3BS we observed numerically all types of solutions in any
vicinity of the initial conditions of general type. This confirms the Poincar\'e hypothesis.

\begin{figure}[!ht]
\centering
\begin{minipage}{8.cm}
\vskip .truecm
\hskip .truecm
\includegraphics[width=\textwidth,natwidth=200,natheight=200]{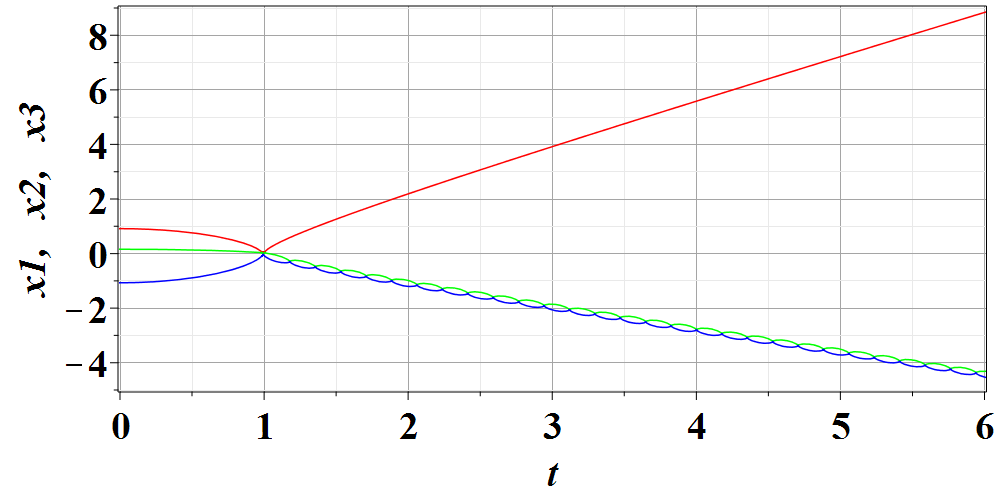}
\vskip .truecm
\hskip .truecm
\includegraphics[width=.5\textwidth,natwidth=200,natheight=200]{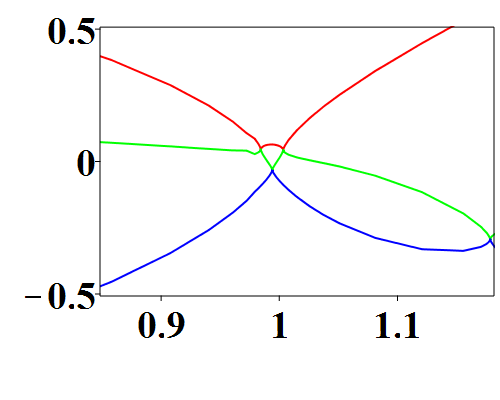}
\end{minipage}
\vskip .truecm
\caption{\small The first example: particle trajectories.}
\label{Fig4}
\end{figure}
\begin{figure}[!ht]
\centering
\begin{minipage}{8.cm}
\vskip .truecm
\hskip .truecm
\includegraphics[width=\textwidth,natwidth=200,natheight=200]{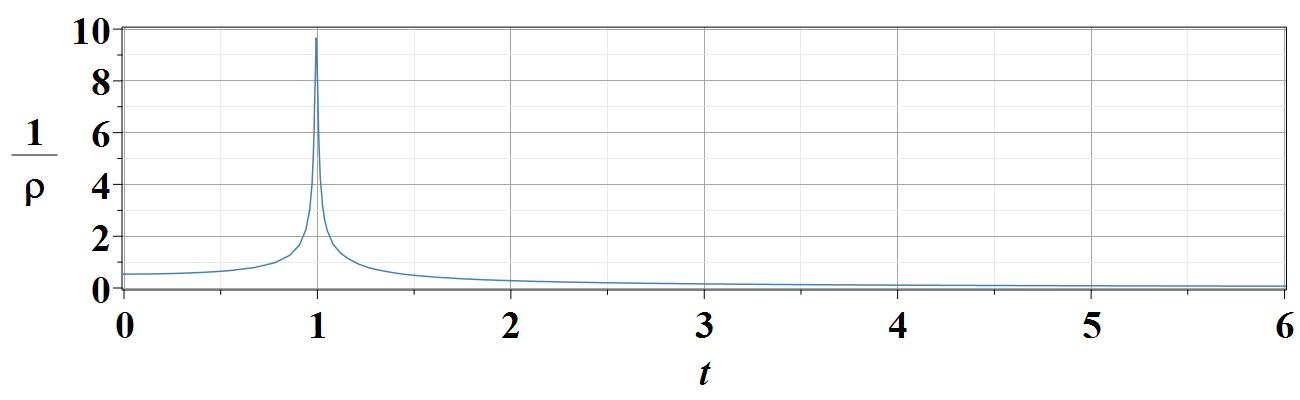}
\vskip .truecm
\hskip .truecm
\includegraphics[width=\textwidth,natwidth=200,natheight=200]{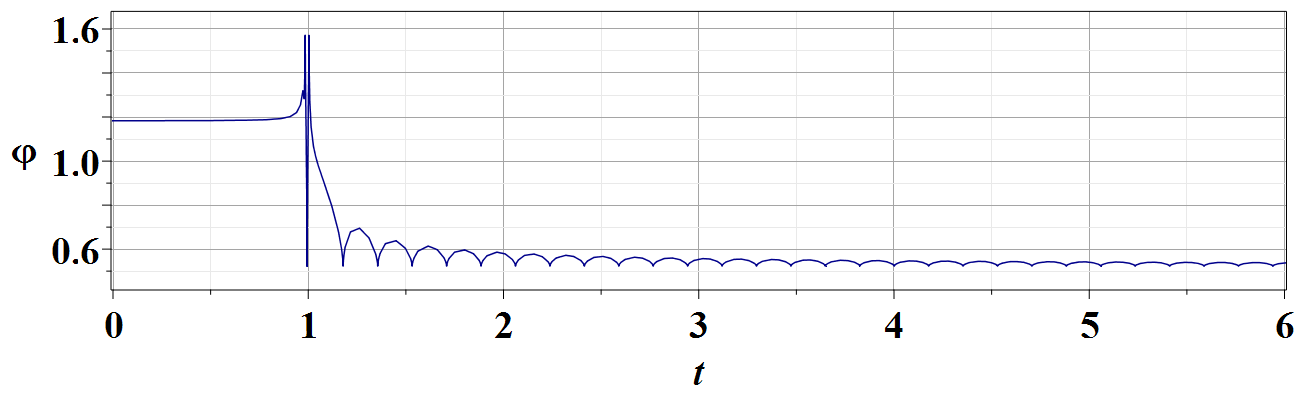}
\end{minipage}
\vskip .truecm
\caption{\small The first example: $\rho(t)^{-1}$ and $\varphi(t)$.}
\label{Fig5}
\end{figure}
\begin{figure}[!ht]
\centering
\begin{minipage}{8.cm}
\vskip .truecm
\hskip .truecm
\includegraphics[width=\textwidth,natwidth=200,natheight=200]{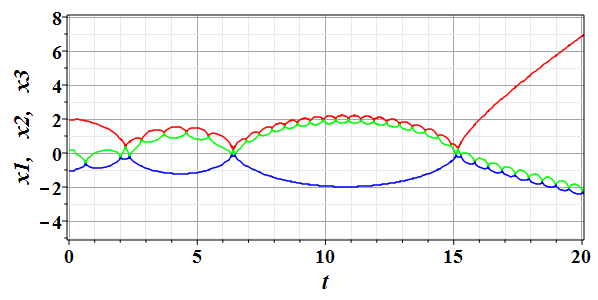}
\end{minipage}
\vskip .truecm
\caption{\small The second example: particle trajectories.}
\label{Fig6}
\end{figure}
\begin{figure}[!ht]
\centering
\begin{minipage}{8.cm}
\vskip .truecm
\hskip .truecm
\includegraphics[width=\textwidth,natwidth=200,natheight=200]{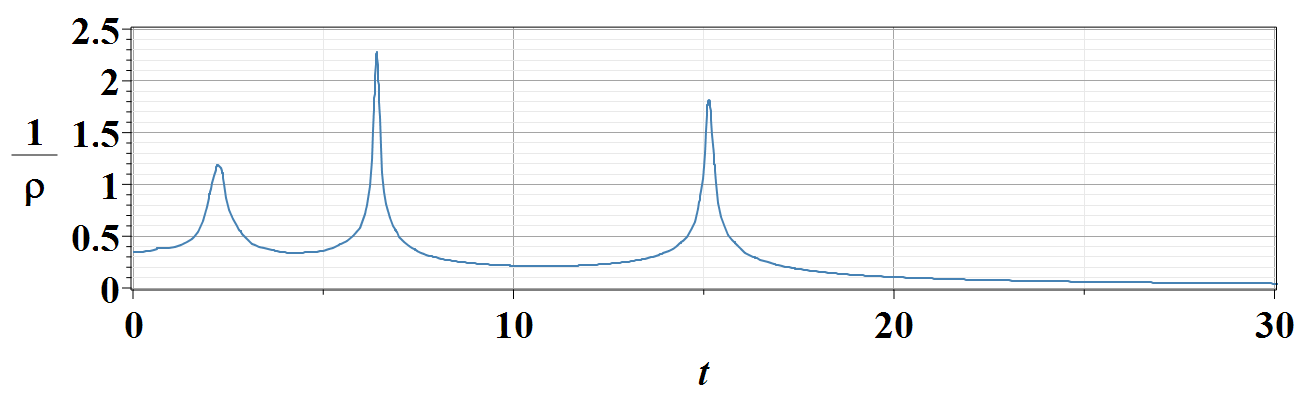}
\vskip .truecm
\hskip .truecm
\includegraphics[width=\textwidth,natwidth=200,natheight=200]{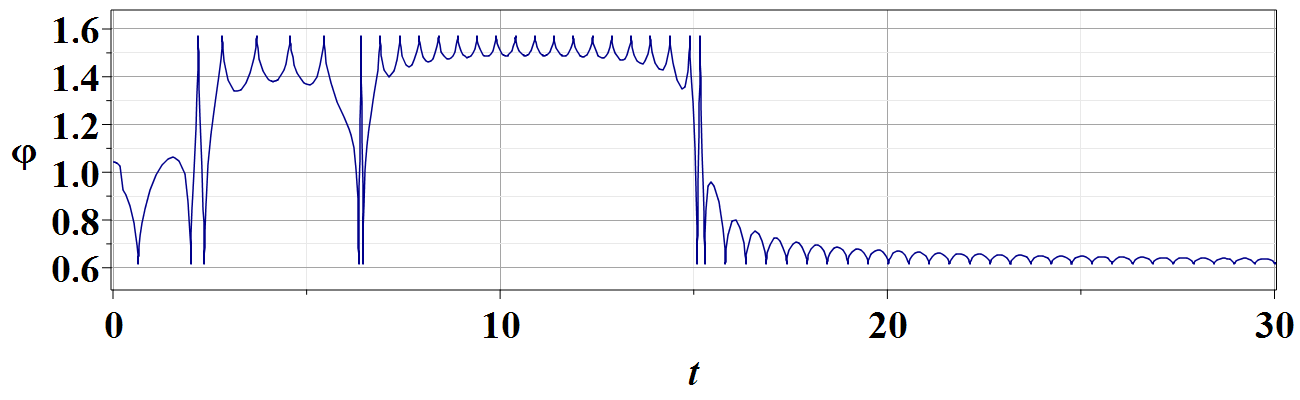}
\end{minipage}
\vskip .truecm
\caption{\small The second example: $\rho(t)^{-1}$ and $\varphi(t)$.}
\label{Fig7}
\end{figure}

We refer the reader to \cite{Fiziev87a,Fiziev87b,Fiziev86,Fiziev87c} for a detailed analytical and numerical discussion of all above statements.
For the purposes of the present Letter we will consider three numerical examples with a proper behaviour to allow decisive conclusions
about the emission of GW in the collinear 3BS.

In the first example (see Figs.\ref{Fig4},\ref{Fig5}), we have equal masses $m_1 = m_2= m_3=1$.
In the second example (see Figs.\ref{Fig6},\ref{Fig7}), we have masses $m_1= m_2=1, m_3=2$.
In both cases the 3BS starts from some positions of the bodies with zero velocities ($\Rightarrow E<0$),
goes through some amount of binary cycles and triple resonances and ends as a binary system
plus a separate particle which goes to infinity.

\begin{figure}[!ht]
\centering
\begin{minipage}{8.cm}
\vskip .truecm
\hskip .truecm
\includegraphics[width=\textwidth,natwidth=200,natheight=200]{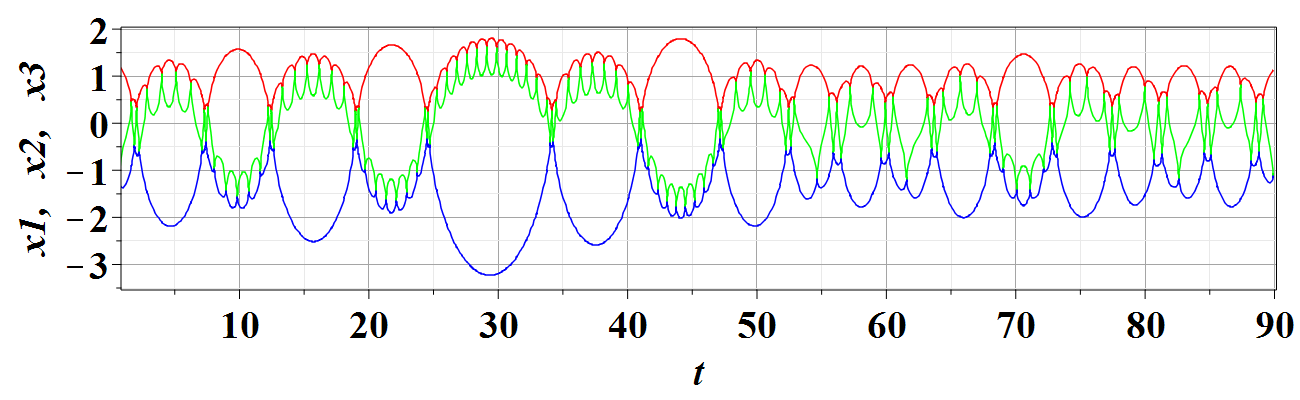}
\end{minipage}
\vskip .truecm
\caption{\small The third example: particle trajectories.}
\label{Fig8}
\end{figure}

\begin{figure}[!ht]
\centering
\begin{minipage}{8.cm}
\vskip .truecm
\hskip .truecm
\includegraphics[width=\textwidth,natwidth=200,natheight=200]{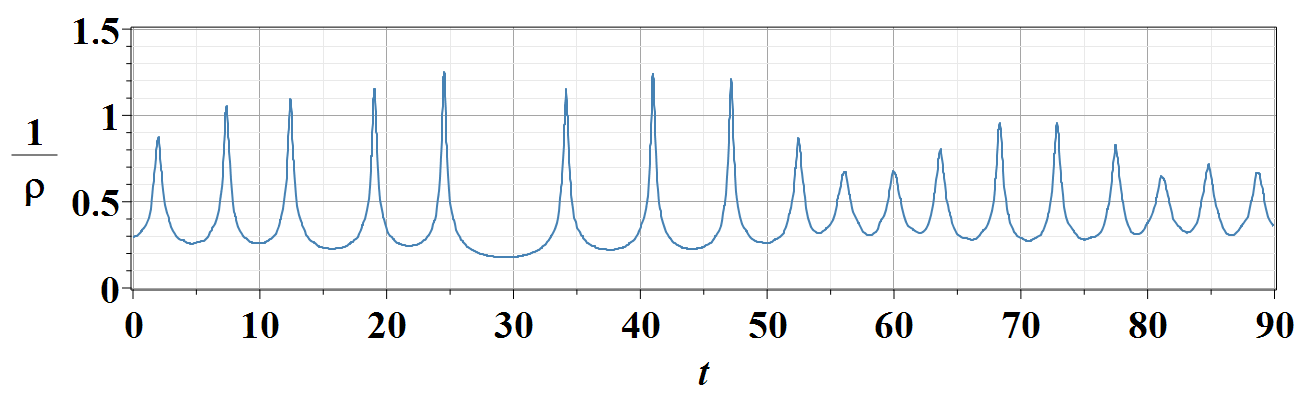}
\vskip .truecm
\hskip .truecm
\includegraphics[width=\textwidth,natwidth=200,natheight=200]{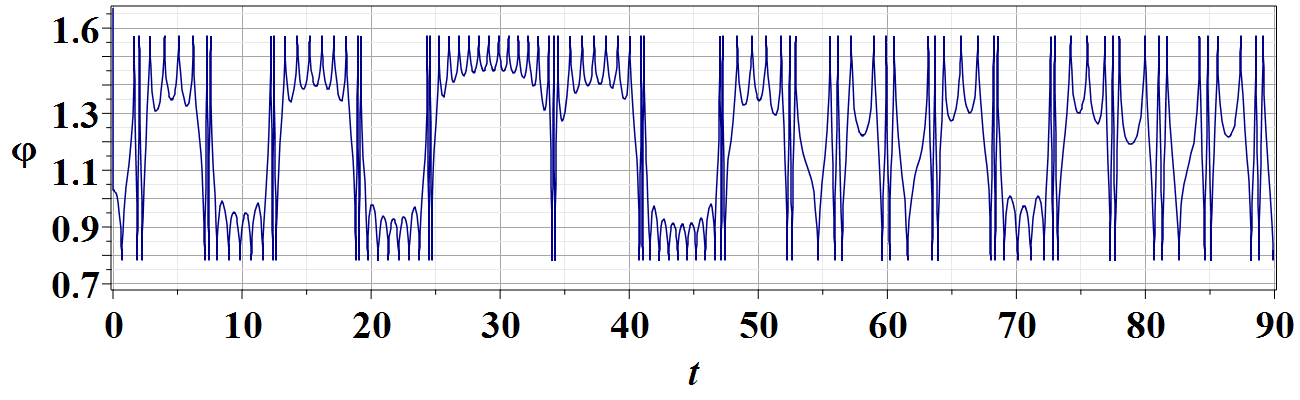}
\end{minipage}
\vskip .truecm
\caption{\small The third example: $\rho(t)^{-1}$ and $\varphi(t)$. }
\label{Fig9}
\end{figure}

In the third example (see Figs.\ref{Fig8},\ref{Fig9}), 3BS with the same start and masses  $m_1=3, m_2=1, m_3=2$ has
a more complicated behaviour and goes trough many different in length binary cycles and  triple resonances,
demonstrating a clear irregular motion.

The same processes go also back in time but are not time-symmetric
when describe recombinations  \cite{Fiziev87a,Fiziev87b,Fiziev86,Fiziev87c}.

\section{Gravitational waves from collinear three body systems}
In the case of GW emission from the collinear 3BS we can choose the axis $Oz$ for GW propagation along it.
Then the only nonzero GW strain  is $h_{+}$\footnote{We have $h_{+}\neq0$ and $h_{\times}\neq 0$
after rotation of the axes $Ox$ and $Oy$ around GW-detector direction $Oz$.}.
In hyperspherical coordinates it acquires the form
\ben
h_{+}={\frac G {r c^4}}{\frac {d^2}{dt^2}}\left(\mu \rho^2\right)={\frac {2G} {r c^4}}\left(2E-V\right).
\la{hp}
\een
Here we used the Lagrange identity \cite{Zeigel71,Arnold06}. The distance from the source to the GW detector is $r$.

If one ignores the back reaction of GW radiation on the 3BS motion, i.e. assuming $E=const$, and ignoring the
unteresting constant component $h_{+,E}={\frac {4 G} {r c^4}} E={\frac{r_s(E/c^2)}{r}}$, one obtains from Eq. \eqref{V} the time-variable part
\ben
h_{+,V}(t)={\frac {r_{s}(\mu) r_{s}(M)} {2 r} }\,{\frac {\alpha} \rho},
\la{hpt}
\een
where $r_s(m)={\frac {2 G m}{c^2}}$ is the Schwarzschild radius of a mass $m$, $c$ is the speed of light, and $\alpha=\alpha(\varphi(t))$, $\rho=\rho(t)$.

An immediate consequence from Eq. \eqref{hpt} and total separation of 2BC from 3BC is the following composition of the corresponding GW-spectra $\widetilde{\alpha}(\omega)$ and  $\widetilde{\rho^{-1}}(\omega)$:
\ben
\widetilde{ h_{+,V}}(\omega)=
{\frac {r_{s}(\mu) r_{s}(M)} {2 r} } \int_{-\infty}^\infty \widetilde{\rho^{-1}}(\omega-\omega^\prime)\widetilde{\alpha}(\omega^\prime)d\omega^\prime.
\la{ar}
\een
Here $\widetilde{f}(\omega)={\frac 1 {2\pi}}\int_{-\infty}^\infty f(t) \exp(-i\omega t) dt$ is
the standard inverse Fourier transform of the function $f(t)$
\footnote{In a small vicinity of 2BC with angle $\varphi_{i,jk}$ we obtain:
$\rho(t) \sim \rho_c=\text{const}$,  $|\Delta \varphi| \sim \left(|\Delta t|/t_c  \right)^{2/3}$,
where $t_c={\frac 2 3}{\frac {\rho_c} {c}}\sqrt{{\frac {\rho_c } {r_s(M)}}{\frac {s_i^2} { c_i}}}$,
$\Delta \varphi=\varphi-\varphi_{i,jk}$, $\Delta t=t-t_i$.
As a result $\alpha(\varphi(t)) \sim \left(|\Delta t|/t_c \right)^{-2/3}$
has an integrable singularity and the spectrum \eqref{ar} is finite everywhere,
although the strain \eqref{hp} grows infinitely in moment $t_i$ of the 2BC.}.

\begin{figure}[!ht]
\centering
\begin{minipage}{8.cm}
\vskip .truecm
\hskip .truecm
\includegraphics[width=\textwidth,natwidth=200,natheight=200]{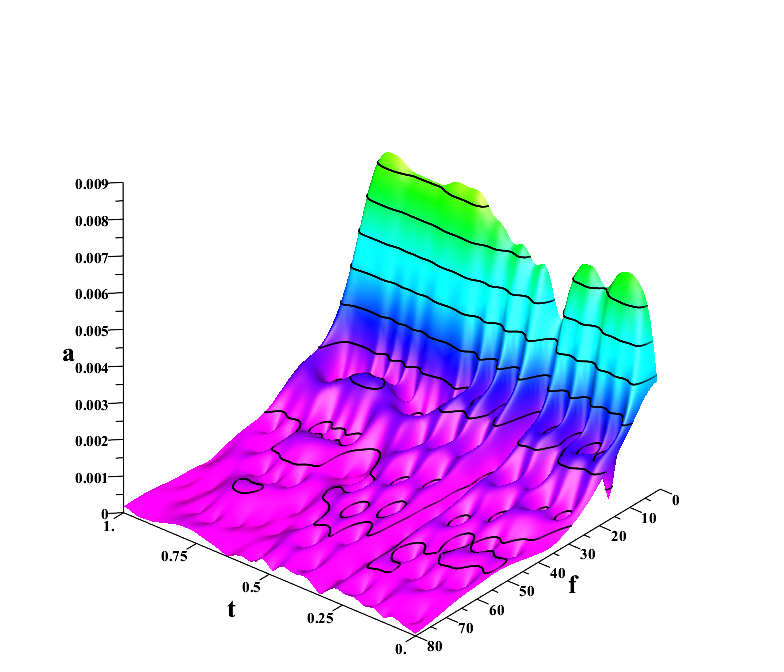}
\vskip -.truecm
\hskip .truecm
\includegraphics[width=\textwidth,natwidth=200,natheight=200]{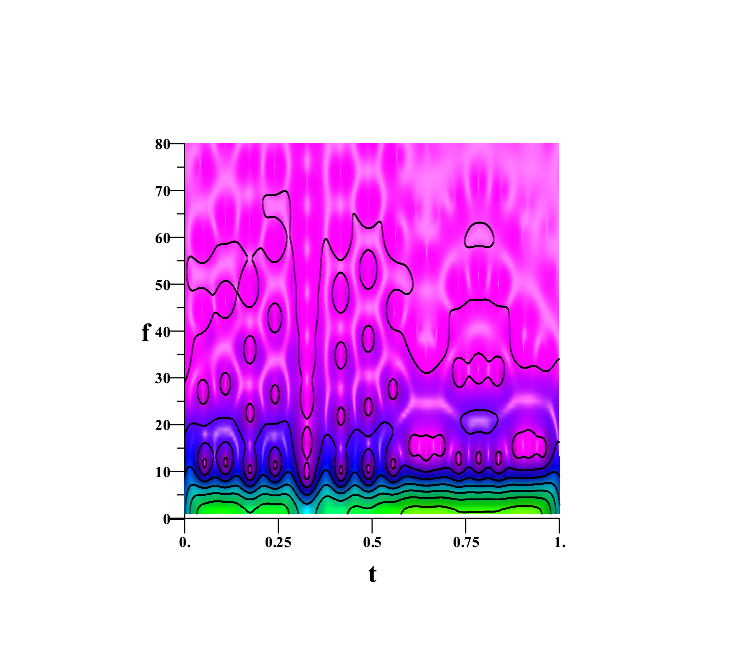}
\end{minipage}
\vskip .truecm
\caption{\small The Gabor transform of $\rho^{-1}(t)$, third example.}
\label{Fig10}
\end{figure}

\begin{figure}[!ht]
\centering
\begin{minipage}{8.cm}
\vskip .truecm
\hskip .truecm
\includegraphics[width=\textwidth,natwidth=200,natheight=200]{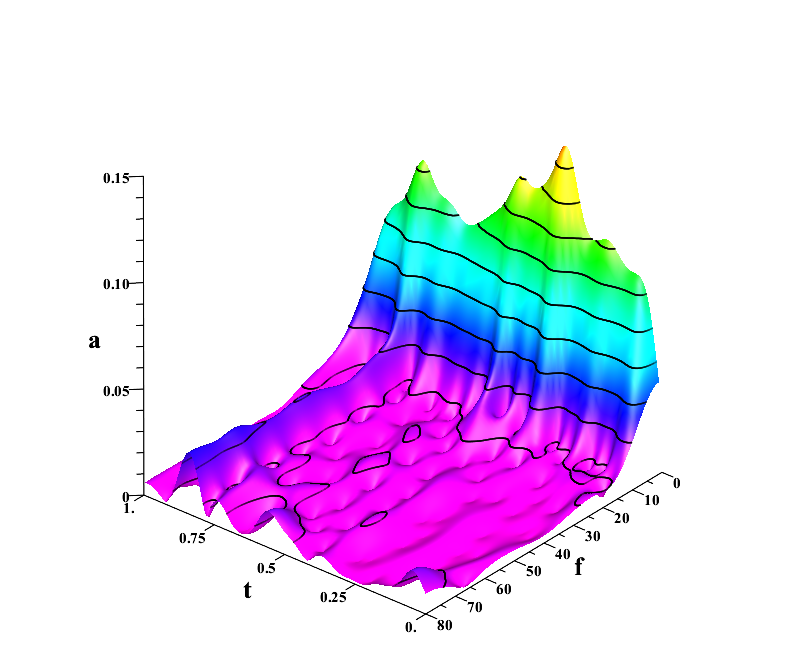}
\vskip .truecm
\hskip .truecm
\includegraphics[width=\textwidth,natwidth=200,natheight=200]{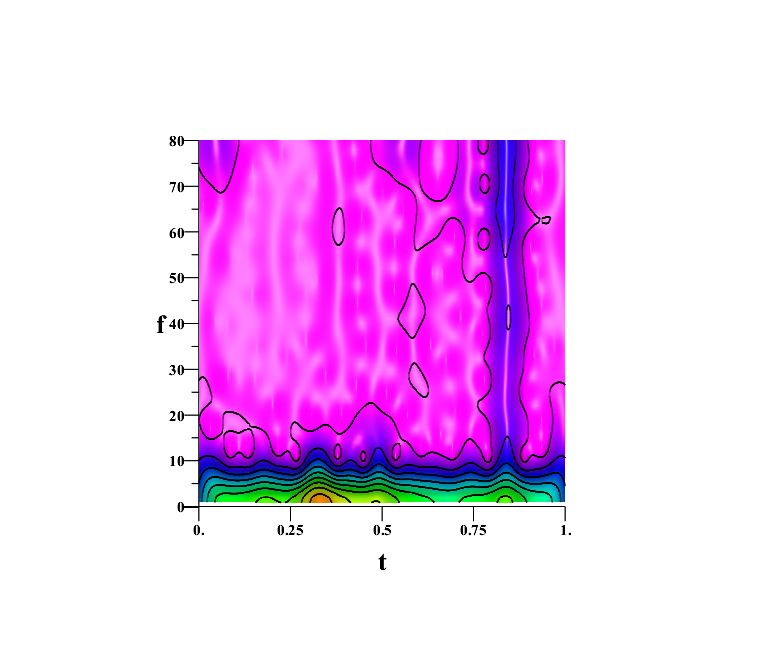}
\end{minipage}
\vskip .truecm
\caption{\small The Gabor transform of $\alpha(t)$, third example.
According to [22], the function $\alpha(t)$ is unbounded but integrable.}
\label{Fig11}
\end{figure}

\begin{figure}[!ht]
\centering
\begin{minipage}{8.cm}
\vskip .0truecm
\hskip .truecm
\includegraphics[width=\textwidth,natwidth=200,natheight=200]{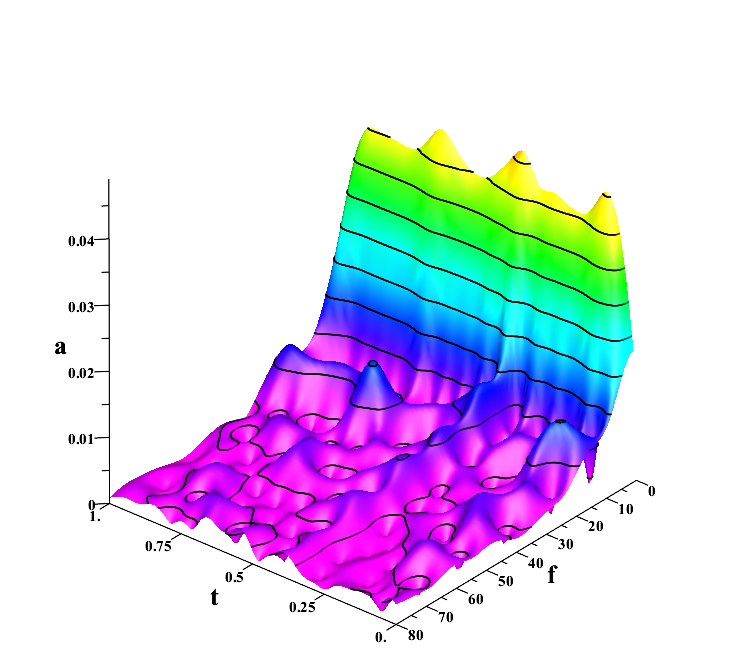}
\vskip .truecm
\hskip .truecm
\includegraphics[width=\textwidth,natwidth=200,natheight=200]{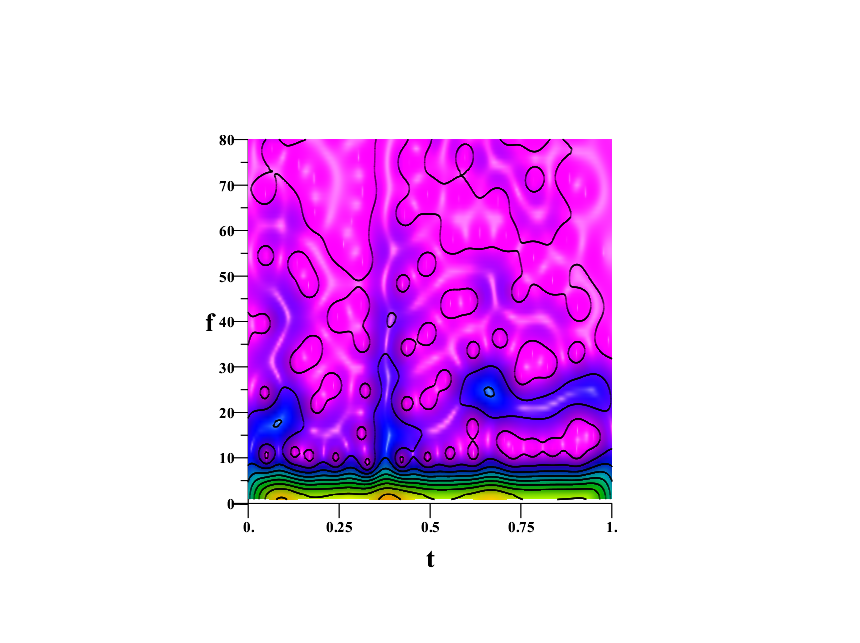}
\end{minipage}
\vskip .truecm
\caption{\small The Gabor transform of $\alpha(t)/\rho(t)$, third example.
The low hills present. This shows that these hills may be a part of the physical signal
describing irregular motion of 3BS.}
\label{Fig12}
\end{figure}

As an illustration of the relations \eqref{hpt} and \eqref{ar} we present
in Figs.\ref{Fig10}-\ref{Fig12}  numerical results for the most interesting third example shown in Figs.\ref{Fig8} and \ref{Fig9}.

Of course, Figs.\ref{Fig10}-\ref{Fig12} do not have a direct relation with Fig.\ref{Fig1}.
However, the comparison clearly shows that the small hills in Fig.\ref{Fig1} most likely
are related with an irregular motion of the source of GW and do not present "a noise of unknown nature".
The most significant physical difference between Fig.\ref{Fig1} and Figs.\ref{Fig10}-\ref{Fig12} is the absence of
chirping frequency in the last ones. According to [19],
it will appear with changes of the 3BS total energy because to the emission of GW.
Hence, for real explanation of the observed GW we need:
1) to consider the back reaction of the emission of GW on the 3BS
motion,
2) to make our consideration relativistic,
3) The next step in the study of GW must be a full numerical-relativity-treatment of  3BS.
The corresponding templates are urgently needed.

This paper was supported by the Sofia University Foundation TCPA, BLTF, JINR, and
Bulgarian Nuclear Regulatory Agency, Grants for 2014, 2015, and 2016.



\begin{thebibliography}{}

\bibitem{LIGO}B. P. Abbott et al.,  Phys. Rev. Lett. {\bf 116} 061102 (2016).

\bibitem{LOGOOPEN} https://losc.ligo.org/events/GW150914/

\bibitem{Thorne87} K. S. Thorne, in Three Hundred Years of Gravitation, eds. S.W. Hawking and W. Israel, Cambridge University Press, Cambridge, England, 1987.

\bibitem{Maggiore} M. Maggiore, {\em Gravitational Waves}, Oxford University Press, 2008.

\bibitem{Buonanno15} A.~Buonanno, B.~Sathyaprakash, {\em Sources of Gravitational Waves: Theory and Observations}, arXiv:1410.7832.

\bibitem{Miller16b} M. C. Miller, Gen. Rel. and Grav. {\bf 48} 95 (2016).

\bibitem{Chiba07} T. Chiba, T. Imai, H. Asada, Mon. Not. R. Astron. Soc. {\bf 377} 269 (2007).

\bibitem{Torigoe09} Y. Torigoe, K. Hattori, H. Asada, Phys. Rev. Lett. {\bf 102} 251101 (2009).

\bibitem{Asada09} H. Asada, Phys. Rev. D 80, 064021 (2009).

\bibitem{Zeigel71} C.L, Zeigel, J.K. Moser {\em Lectures on Celestial Mechanics}, Springer, 1971.

\bibitem{Arnold06} V.I. Arnold, V.V. Kozlov, A.I. Neishtadt, {\em Mathematical Aspects of Classical and Celestial Mechanics}, Springer, 2006.

\bibitem{Dmitracinovic} V. Dmitrasinovic, M. Suvakov, A. Hudomal,  Phys. Rev. Lett. {\bf 113} 101102 (2014).

\bibitem{McGehee74} R. McGehee, lnventiones math. {\bf 27} 191 (1974).

\bibitem{Fiziev87a} P.P. Fiziev, {\em Investigation of integrability of classical three body problem}, PhD Thesis, JINR, Dubna, 1987.
DOI: 10.13140/RG.2.2.13686.40008. https://www.researchgate.net/publication/307208487 \_Investigation\_of\_the\_integrability\_of\_the\_classical\_three-body\_problem?showFulltext=1 \&linkId=57c4658d08ae32a03dad4111

\bibitem{Fiziev87b} P.P. Fiziev, Ts.Ya. Fizieva, Few-Body Syst. {\bf 2} 71 (1987)

\bibitem{Fiziev86} P.P. Fiziev, Letters in Math. Phys. {\bf 12}  267 (1986).

\bibitem{Fiziev87c} P.P. Fiziev, T.M. Mishonov, Ts.Ya. Fizieva,  preprint P2-87-737, JINR, Dubna (1987).
http://www-lib.kek.jp/cgi-bin/kiss\_prepri.v8?KN=198801219\&OF=4.

\end{thebibliography}
\end{document}